\documentclass[conference]{IEEEtran}
\IEEEoverridecommandlockouts
\usepackage{cite}
\usepackage{stfloats}
\usepackage{amsmath,amssymb,amsfonts}
\usepackage{algorithm}               
\usepackage{algorithmic}             
\usepackage{graphicx}
\usepackage{textcomp}
\usepackage{xcolor}
\usepackage{setspace}

\def\BibTeX{{\rm B\kern-.05em{\sc i\kern-.025em b}\kern-.08em
    T\kern-.1667em\lower.7ex\hbox{E}\kern-.125emX}}
\begin{document}

\title{Bounding Queue Length Violation Probability of Joint Channel and Buffer Aware Transmission
}

\author{\IEEEauthorblockN{Lintao Li\IEEEauthorrefmark{1}\IEEEauthorrefmark{2},  Wei Chen\IEEEauthorrefmark{1}\IEEEauthorrefmark{2}, \emph{Senior Member, IEEE}, and Khaled B. Letaief\IEEEauthorrefmark{3}\IEEEauthorrefmark{4}, \emph{Fellow, IEEE}}
	\IEEEauthorblockA{\IEEEauthorrefmark{1}Department of Electronic Engineering, Tsinghua University, Beijing, 100084, CHINA}
	\IEEEauthorblockA{\IEEEauthorrefmark{2}Beijing National Research Center for Information Science and Technology (BNRist)}
	\IEEEauthorblockA{\IEEEauthorrefmark{3}School of Engineering, Hong Kong University of Science and Technology, Hong Kong}
	\IEEEauthorblockA{\IEEEauthorrefmark{4}Peng Cheng Laboratory, Shenzhen, 518066, CHINA}
	\IEEEauthorblockA{Email: llt20@mails.tsinghua.edu.cn, \{eewchen@gmail.com, wchen@tsinghua.edu.cn\}, eekhaled@ust.hk }}

\maketitle

\begin{abstract}
Queue length violation probability, i.e., the tail distribution of the queue length, is a widely used statistical quality-of-service (QoS) metric in wireless communications. Characterizing and optimizing the queue length violation probability have great significance in time sensitive networking (TSN) and ultra reliable and low-latency communications (URLLC). However, it still remains an open problem. In this paper, we put our focus on the analysis of the tail distribution of the queue length from the perspective of cross-layer design in wireless link transmission. We find that, under the finite average power consumption constraint, the queue length violation probability can achieve zero with diversity gains, while it can have a linear-decay-rate exponent according to large deviation theory (LDT) with limited receiver sensitivity. Besides, we find that the arbitrary-decay-rate queue length tail distribution with the finite average power consumption exists in the Rayleigh fading channel. Then, we generalize the sufficient conditions for the communication system belonging to these three scenarios, respectively. Moreover, we apply the above results to analyze the wireless link transmission in the Nakagami-m fading channel. Numerical results with approximation validate our analysis.

\end{abstract}

\section{Introduction}
In the fifth-generation (5G) communication
networks, tremendous demand has been raised on the higher transmission rate and communication reliability. To achieve these goals, ultra reliable and low-latency communications (URLLC) attracts many researchers to work on it. Moreover, the extreme URLLC scenarios in the envision of the six-generation (6G) communication networks also require more efforts on ensuring the quality-of-service (QoS) in communication systems \cite{6G}. In recent years, an emerging concept of time sensitive networking (TSN) also put requirements on the delay performance \cite{tsn}. Therefore, it is essential to characterize and improve the QoS performance in communication systems.

To reduce the transmission latency and power consumption, cross-layer design has been adopted in many works, which combined the queue and channel states to make decisions on transmission rate or power. In our previous works, the optimal control policies and delay-power tradeoffs were obtained by using constrained Markov decision process (CMDP) in adaptive transmission \cite{chen}, bursty random arrivals \cite{wang}, and Markovian arrivals systems \cite{zhao}. Besides, work \cite{yang} proposed the cross-layer power allocation scheme to minimize the overall delay in multi-access channels. These works give us a way to obtain the optimal cross-layer control policies in different communication systems.

Having obtained the transmission policies, we then need to analyze the QoS performance in communication systems by adopting these policies. In wireless link transmission, our focus is mainly on the statistical QoS performance, i.e., length or delay violation probabilities. One of the most widely used analysis tools is the large deviation theory (LDT),  which gives a closed-form approximation for length violation probability. Based on LDT, in \cite{ec}, effective capacity was proposed to characterize the service ability of wireless communication systems with QoS constraints. The maximum throughput constrained by the QoS requirements was studied in \cite{ecqos} using a large deviations framework. The authors of \cite{spldt} investigated different downlink transmission scheduling schemes in a cellular network under QoS constraints. Both \cite{ecqos} and \cite{spldt} derived the exponent of the length violation probability by finding the most likely path to overflow. In \cite{lya}, the authors combined sample-path LDT with Lyapunov functions to generalize the above analysis. However, since the analysis is complex, only the max-queue policy with on-off channels was analyzed in detail. Besides, in these works, the authors all assumed that the condition of LDT is satisfied by the policy, i.e., the queue length violation probability has a linear-decay-rate exponent with the queue length. However, not all policies meet that condition, which means the analysis based on LDT has its limitation. Thus, characterizing the achievable decay rate of the queue length violation probability is important. It has significance in evaluating the optimality of the proposed policies.

In this paper, we put our focus on the analysis of the length violation probability, i.e., the queue length tail distribution, from the perspective of cross-layer design in the single-input single-output (SISO) system. We find that, in the Rayleigh fading channel, under the finite average power consumption constraint, the queue length violation probability can be equal to zero with diversity gains, while it can have a linear-decay-rate exponent according to LDT with limited receiver sensitivity. Besides, we find that the arbitrary-decay-rate queue length tail distribution exists in the Rayleigh fading channel with the finite average power consumption. Based on the above findings, we divide the wireless link transmission systems into three scenarios according to the decay rate of the queue length tail distribution with the finite average power consumption. We then generalize the sufficient conditions for the communication system belonging to these three scenarios, respectively. Then, we use the above results to conduct analysis in the Nakagami-m fading channel.

Through the analysis in this paper, we prove the existence of the queue length tail distribution with arbitrary-decay-rate exponents under the finite average power consumption constraint. This result indicates that LDT is not enough for analyzing the queue length violation probability in wireless link transmission systems.
Besides, by characterizing the achievable decay rate of the queue length violation probability and their corresponding sufficient conditions in different scenarios, we give guidance on designing a high-performance wireless transmission policy.

Throughout this paper, $\mathbb{E}\{x\}$ denotes the expectation of random variable $x$. $\mathbb{R}$ and $\mathbb{C}$ denote the set of real numbers and complex numbers, respectively.

\section{System Model}
Consider a wireless link transmission system, in which the transmitter sends data to the receiver via a time-varying channel as shown in Fig. 1. In this section, we will introduce the physical layer and network layer model, respectively.

\subsection{Physical Layer}
In this system, time is divided into timeslots with equal length $T_0$. The data will be transmitted over a block fading channel, in which the channel coefficient varies in an $i.i.d.$ manner across time slots. Let $h[n]\in \mathbb{C}$ denote the channel coefficient in the time slot $n$. Then, for the input $x[n] \in \mathbb{C}$ in time slot $n$, the corresponding output $y[n] \in \mathbb{C}$ is given by
\begin{equation}
y[n]=h[n]x[n]+z[n], \label{eq1}
\end{equation}
where $z[n] \in \mathbb{C}$ is the additive Gaussian noise with power $\sigma^2$. We assume that the channel state information (CSI) is available at both the transmitter and the receiver.  The adaptive transmitter determines the transmission rate in each time slot. According to Shannon's formula, with the bandwidth $B$, the power consumption $P[n]$ corresponding to the transmission rate $R[n]$ (in nats/s) is given by
\begin{equation}
P[n]=\frac{\sigma^2}{|h[n]|^2}\bigg(e^{\big(\frac{R[n]}{B}\big)}-1\bigg). \label{eq2}
\end{equation}

\begin{figure}[t]
	\centerline{\includegraphics[width=8.5cm]{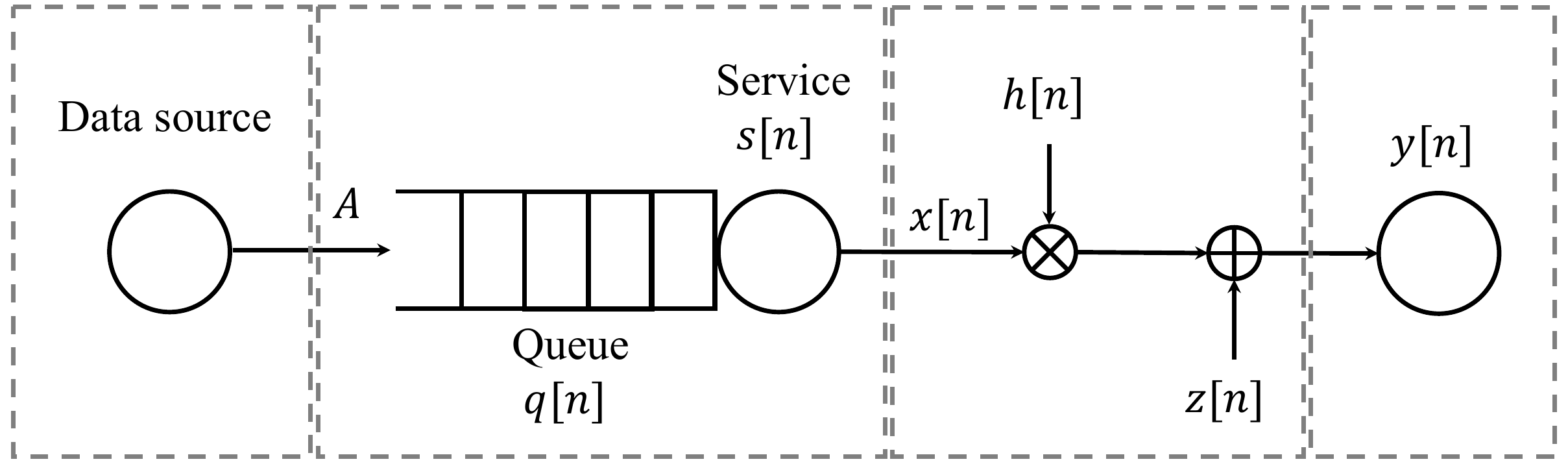}}
	\caption{System Model.}
	\label{sys}
\end{figure}

\subsection{Network Layer}
In this paper, we assume that the size of arrival data is deterministic in each time slot, i.e., $A$ nats data arrive in each time slot. Besides, we denote $s[n]$ as the size of transmission data at the end of time slot $n$, which is equal to $T_0R[n]$ in this system. At the transmitter, an infinite-length buffer is equipped to backlog the arrival data. Let $q[n]$ denote the length of the queue at the beginning of time slot $n$. Consequently, the queue length is updated as
\begin{equation}
q[n+1]=\max \big \{ q[n]+A-s[n],0 \big \} \label{eq3}
\end{equation}
To keep the stability of the queue, we assume that $A<\mathbb{E}\{s[n]\}$.

 For a cross-layer scheduling policy, the transmitter determines $s[n]$ according to $q[n]$ and $h[n]$. Let $g(x)$ denote the probability density function ($p.d.f.$) of $|h[n]|^2$, $u(x)$ denote the $p.d.f.$ of $q[n]$, and   $f_{q,h}^s$ denote the probabilistic cross-layer scheduling policy, which is the conditional $p.d.f.$ for $s[n]=s$ under the condition that $|h[n]|^2=h$ and $q[n]=q$. Then, the average power consumption of this policy is given by
\begin{equation}
P_{\rm avg}=\int_{0}^{\infty}\int_{0}^{\infty}\int_{0}^{q+A}\!\!\frac{\sigma^2 }{h}\big(e^{\frac{s}{T_0B}}\!-\!1\big) f_{q,h}^{s} g(h)u(q) ds\, dh\, dq. \label{eq4}
\end{equation}

Given a queue length threshold $q_{\rm th}$, the queue length violation probability $\varepsilon(q_{\rm th})$, i.e., the tail distribution of the queue length, is defined as 
\begin{equation}
	\varepsilon(q_{\rm th})=\lim_{T\to \infty}\frac{1}{T} \sum_{t=1}^T\mathbb{I}\{q[t]>q_{\rm th}\},
\end{equation}
where $\mathbb{I}\{\cdot\}$ is the indicator function.

\section{Tail Distribution Analysis of the Queue Length}
In this section, we will present the relationships between the channel distribution and the tail distribution of the queue length adopting control policies with the finite average power consumption. Specifically, we first give the sufficient conditions on the channel distribution for achieving zero queue length violation probability with finite $P_{\rm avg}$, which is called Scenario 1 in the following. Besides, we also present sufficient conditions on the channel distribution for achieving the tail distributions of the queue length with the linear-decay-rate and arbitrary-decay-rate exponent given finite $P_{\rm avg}$, which are called Scenario 2 and Scenario 3, respectively.
For simplicity, we assume $A=1$ and $T_0B=1$ in this section.

\subsection{Non-Length-Violation Scenario}
For Scenario 1, to achieve zero queue length violation probability for arbitrary $q_{\rm th}$, one intuitive way is that the transmitter sends all the arrival data in each time slot, i.e., $s[n]=1$. However, this policy may result in infinite power consumption. For example, in \cite{inversion}, the author proved that the channel inversion policy has an infinite average power consumption in Rayleigh fading channels. We can find that, with diversity gains, the wireless link transmission system in the Rayleigh fading channel belongs to Scenario 1.  In Lemma 1, we give the sufficient condition on the channel distribution for achieving zero queue length violation probability with finite $P_{\rm avg}$.

\textbf{Lemma 1.} The sufficient condition for $\varepsilon(q_{\rm th})=0$, $q_{\rm th}>1$, with finite $P_{\rm avg}$, is given by
\begin{equation}
\lim_{x\to 0^+}g(x)=0, \label{eq7}
\end{equation}
and 
\begin{equation}
\lim_{x\to 0^+}g'(x)<\infty, \label{eq8}
\end{equation}
where $g'(x)$ is the derivative of $g(x)$.

\begin{IEEEproof}
Based on Eq. \eqref{eq4}, to transmit all the arrival data in each time slot, the average power consumption can be expressed as 
\begin{equation}
P_{\rm avg}=(e-1)\sigma^2\int_{0}^{\infty}\frac{g(x)}{x}dx. \label{eq9}
\end{equation}
To keep this integral finite, the key point is to avoid the influence of $\lim_{x\to0^+}\frac{1}{x}= \infty$. One sufficient condition is to control the integrand not approaching infinity when $x$ approaches 0, i.e. $\lim_{x\to 0^+}\frac{g(x)}{x}<\infty$.
From Eq. \eqref{eq9}, we can find that if $P_{\rm avg}$ is finite, then $\lim_{x\to 0^+}g(x)=0$.  Since $\lim_{x\to 0^+}g(x)=0$, $g(x)$ can then be expanded as 
\begin{equation}
g(x)=g'(x)x+\mathcal{O}(x), \label{eq10}
\end{equation}
where $\mathcal{O}(x)$ denote the higher-order infinitesimal of $x$. Based on Eq. \eqref{eq10}, now $\lim_{x\to 0^+}\frac{g(x)}{x}<\infty$ is equivalent to $\lim_{x\to 0^+}g'(x)<\infty$. Thus, if Eqs. \eqref{eq7} and \eqref{eq8}  hold, $P_{\rm avg}$ must be finite. 
\end{IEEEproof}
For $g(x)$ satisfying Lemma 1, we can conclude that with this channel the length violation probability for arbitrary $q_{\rm th}>1$ is zero with finite power consumption. Thus, the communication system with $g(x)$ satisfying Lemma 1 belongs to Scenario 1.

\subsection{Tail Distributions of the Queue Length with Linear-Decay-Rate and Arbitrary-Decay-Rate Exponents}
According to the proof of Lemma 1, for $g(x)$ not satisfying Lemma 1, we can divide them into two kinds. One kind can guarantee at least the tail distribution of queue length with a linear-decay exponent given finite $P_{\rm avg}$, i.e., it belongs to Scenario 2. Another kind can make sure that the tail distribution of queue length has arbitrary-decay-rate exponent given finite $P_{\rm avg}$, i.e., it belongs to Scenario 3. 

For the fixed power policy, i.e., $P[n]=P$, according to effective capacity \cite{ec}, with the help of LDT, the length violation probability satisfies
\begin{equation}
	\lim_{q_{\rm th}\to \infty}\frac{\ln \big( \varepsilon(q_{\rm th}) \big)}{q_{\rm th}}=-\theta, \label{eq5}
\end{equation} 
where $\theta$ is called the decay exponent. It can be obtained by solving 
\begin{equation}
	-\lim_{n \to \infty} \frac{1}{n \theta} \ln \mathbb{E}\bigg \{e^{-\theta BT_0\ln\big(1+\frac{|h[n]|^2P}{\sigma^2}\big)}\bigg\}=A. \label{eq6}
\end{equation} 
 Above results in \cite{ec} shows that the fixed power policy has a linear-decay-rate exponent $-\theta q_{\rm th}$ along with $q_{\rm th}$. Lemma 2 gives an intuitive condition for $g(x)$ belongs to Scenario 2 instead of Scenario 1.

\textbf{Lemma 2}. If $g(x)$ satisfies
\begin{equation}
\lim_{\epsilon\to 0^+}\int_{0}^{\epsilon} g(x) \neq 0, \label{eq11}
\end{equation}
then it belongs to Scenario 2.

Eq. \eqref{eq11} indicates that $g(x)$ has an impulse on $x=0$, which means there are probabilities that $|h[n]|^2=0$ occurs. This is possible because the receiver can not work with the limited receiver sensitivity if $|h[n]|^2$ is too small. When it occurs, the transmitter can not send any data on that time slot with the finite power consumption, which is equivalent to $|h[n]|^2=0$. This indicates that $g(x)$ satisfying Eq. \eqref{eq11} does not belong to Scenario 1. Whether it belongs to Scenario 3 is not sure. However, we can adopt the fixed power policy to at least guarantee that the queue length tail distribution has a linear-decay-rate exponent with finite $P_{\rm avg}$, i.e., it belongs to Scenario 2. 

Before conducting an analysis on Scenario 3, we first give its definition as follows.
 
\textbf{Definition 1}. In Scenario 3, the queue length tail distribution satisfies the following relationship with finite $P_{\rm avg}$:
\begin{equation}
\lim_{q_{\rm th} \to \infty}\frac{\ln \big(\varepsilon(q_{\rm th})\big)}{c_m(q_{\rm th})}=V_m,
\end{equation}
where $V_m \in [0,\infty)$ is a constant, and $\{c_m(q_{\rm th}),m=0,1,2\cdots\}=\{-q_{\rm th}, -\exp(q_{\rm th}),-\exp\big (\exp(q_{\rm th})\big),\cdots\}$.

For Scenario 3, we first conceive one transmission policy, in which the queue length and transmission rate can only be integer. Let $h_{\rm th}^{k}$ denote the threshold of the channel power gain with queue length $k$. When $q[n]=k$ and $|h[n]|^2<h_{\rm th}^{k}$, the transmitter will not send any data at time slot $n$. The policy can be expressed as 
\begin{align}
s[n]=\begin{cases}
1, \quad \text{If $h[n]\geq h_{\rm th}^{0}$, $k = 0 $},\\
2, \quad \text{If $h[n]\geq h_{\rm th}^{k}$, $k \geq 1$},\\
0, \quad \text{If $h[n]< h_{\rm th}^{k}$, $k \geq 0$}.\label{eq12}
\end{cases}
\end{align}

 Let $p_k$ denote the probability of not sending any data when $q[n]=k$. The relationship between $p_k$ and $h_{\rm th}^k$ is given by
 \begin{equation}
 p_k=\int_{0}^{h_{\rm th}^{k}}g(x) dx. \label{pk}
 \end{equation}
 Thus, we can control this policy by presenting different sequences of $p_k$.
 
Then, we will prove that, by adopting this policy, the tail distribution of queue length can have arbitrary-decay-rate exponents with finite $P_{\rm avg}$ in the Rayleigh fading channel. In Theorem 1, we first give the sufficient condition for that the above transmission policy has finite $P_{\rm avg}$ in the Rayleigh fading channel.

\newcounter{TempEqCnt}
\setcounter{TempEqCnt}{\value{equation}} 
\setcounter{equation}{22} 
\begin{figure*}[hb]
	\normalsize \hrulefill
	\begin{equation}
		P_{\rm avg}=\Bigg(1+\sum_{k=1}^{\infty}\prod_{m=0}^{k-1}\bigg(\frac{p_m}{\mu_{m+1}}\bigg)\Bigg)^{-1}  \Bigg( C_0+(e^2-1)\sum_{k=1}^{\infty}\bigg(\prod_{m=0}^{k-1}\frac{p_m}{\mu_{m+1}}\bigg)    \int_{-\ln(1-p_k)}^{\infty}\frac{e^{-x}}{x} dx \Bigg),\label{pave}
	\end{equation}
	\setcounter{equation}{\value{TempEqCnt}}
	\setcounter{equation}{24} 
	\begin{align}
		P_{\rm avg}&=\pi_0(e^2-1)\sum_{k=1}^{\infty}\bigg(\prod_{m=0}^{k-1}\frac{p_m}{\mu_{m+1}}\bigg)\int_{-\ln(1-p_k)}^{\infty}\frac{e^{-x}}{x} dx+\pi_0 C_0 \nonumber \\
		&<\pi_0(e^2-1)\sum_{k=1}^{\infty}\bigg(\prod_{m=0}^{k-1}\frac{p_m}{\mu_{m+1}}\bigg)\mu_k\ln \bigg(1-\frac{1}{\ln(1-p_k)}\bigg)+\pi_0 C_0. \label{thm32}
	\end{align}
\end{figure*}
\setcounter{equation}{\value{TempEqCnt}}

\textbf{Theorem 1}. For a Rayleigh fading channel with $\mathbb{E}\{|h[n]|^2\}=1$, a sufficient condition for that the above transmission policy has finite $P_{\rm avg}$ is given by
\begin{equation}
\lim_{k \to \infty}\frac{p_k \ln(1+\frac{1}{p_{k+1}})}{\mu_{k}\ln(1+\frac{1}{p_k})}<1,\label{eq13}
\end{equation}
where $\mu_k=1-p_k$.
\begin{IEEEproof}
	We start this proof by obtaining the steady state probability of the queue length with the transmission policy. Let $\pi_k$ denote the steady state probability of queue length being equal to $k$, $k=0,1,\cdots$. To obtain $\pi_k$, we present the steady state equation and normalization condition, which are given by
	\begin{align}
	\begin{cases}
	p_0\pi_0=\mu_1\pi_1,\\
	(\mu_k+p_k)\pi_k = p_{k-1}\pi_{k-1} + \mu_{k+1}\pi_{k+1},\,\, \text{$k=1,2,\cdots$}\\
	\sum_{k=0}^{\infty}\pi_k=1.
	\end{cases}
	\end{align}
	By substituting the steady state equation into the normalization condition, we obtain $\pi_0$ as 
	\begin{equation}
	\pi_0=\Bigg(1+\sum_{k=1}^{\infty}\prod_{m=0}^{k-1}\bigg(\frac{p_m}{\mu_{m+1}}\bigg)\Bigg)^{-1}.\label{pi0}
	\end{equation}
	 Besides, from the steady state equation, we can obtain the local equilibrium equation as
	\begin{equation}
	\mu_k\pi_k=p_{k-1}\pi_{k-1}, \quad \text{$k=1,2,\cdots$}.\label{localeq}
	\end{equation}
	By substituting Eq. \eqref{pi0} into Eq. \eqref{localeq}, $\pi_k$, $k=1,\cdots$, is given by
	\begin{equation}
	\pi_k=\pi_0\prod_{m=0}^{k-1}\bigg(\frac{p_m}{\mu_{m+1}}\bigg),\label{pi}
	\end{equation}
	For a Rayleigh fading channel with $\mathbb{E}\{|h[n]|^2\}=1$, Eq. \eqref{pk} can be simplified as
	\begin{equation}
	p_k=\int_{0}^{h_{\rm th}^{k}}e^{-x}dx.\label{theq}
	\end{equation}
	From Eq. \eqref{theq}, we have $h_{\rm th}^{k}=-\ln(1-p_k)$. Let $C_k$ denote the average power consumption with the queue length $k$, which is given by
	\begin{align}
	C_k=
	\begin{cases}
	(e^2-1)\int_{-\ln(1-p_k)}^{\infty}\frac{e^{-x}}{x}dx, \quad &\text{$k=1,2,\cdots$ ,}\\
	(e-1)\int_{-\ln(1-p_0)}^{\infty}\frac{e^{-x}}{x}dx, \quad &\text{$k=0$}.
	\end{cases}\label{ck}
	\end{align}
	Then, by combining Eqs. \eqref{pi}, \eqref{pi0}, and \eqref{ck}, we obtain Eq. \eqref{pave}.

	According to \cite{e1}, the function $E_1(x)=\int_{x}^{\infty}\frac{e^{-t}}{t} dt$ has the property that 
	\setcounter{equation}{23}
	\begin{equation}
	E_1(x)<e^{-x}\ln(1+\frac{1}{x}), \quad x>0 \label{e1x}
	\end{equation}
	By combining Eq. \eqref{e1x} with Eq. \eqref{pave}, we obtain Eq. \eqref{thm32}.

	The Taylor's expansion of $\ln(1-p_k)$ can be expressed as $-p_k-\frac{1}{2}p_k^2+O(p_k^2)$. Therefore, we obtain that $\ln(1-p_i)<-p_i$, which indicates $\ln(1-\frac{1}{\ln(1-p_k)})<\ln(1+\frac{1}{p_k})$. Then, Eq. \eqref{thm32} can be further expressed as
	\setcounter{equation}{25}
	\begin{equation}
	P_{\rm avg}<\pi_0(e^2-1)\sum_{k=1}^{\infty}\bigg(\prod_{m=0}^{k-1}\frac{p_m}{\mu_{m+1}}\bigg)\mu_k\ln(1+\frac{1}{p_k})+\pi_0 C_0. \label{thm33}
	\end{equation}
	Therefore, if the sequence in Eq. \eqref{thm33} is convergent, the average power of this policy is finite. According to D’Alembert’s ratio test \cite{rudin}, we can obtain a sufficient condition of the finite average power consumption as shown in Eq. \eqref{eq13}.
\end{IEEEproof}

\textbf{Corollary 1.} For decreasing sequences $\{p_k, k=0,1, \cdots\}$ with $\lim_{k \to \infty} p_k=0$, a sufficient condition for the above transmission policy having finite $P_{\rm avg}$ can be simplified as
\begin{equation}
\lim_{k \to \infty}\frac{p_k \ln(p_{k+1})}{\mu_{k}\ln(p_k)}<1.\label{coro1}
\end{equation}
\begin{IEEEproof}
	If $p_k$ approaches 0, then $\ln(1+\frac{1}{p_k})\approx \ln(\frac{1}{p_k})$. Combined with Eq. \eqref{eq13}, the sufficient condition is simplified as Eq. \eqref{coro1}.
\end{IEEEproof}

Based on Theorem 1 and Corollary 1, we give the conclusion that the tail distribution of queue length can have an arbitrary-decay-rate exponent in Rayleigh fading channels with finite $P_{\rm avg}$. We summary it in Theorem 2.

\textbf{Theorem 2}. By adopting the above transmission policy, the tail distribution of queue length has an arbitrary-decay-rate exponent with finite $P_{\rm avg}$ in Rayleigh fading channels. The decay rate is dominated by $\sum_{k=q_{\rm th}+1}^{\infty} \prod_{m=0}^{k-1}p_m $.

\begin{IEEEproof}
According to Eqs. \eqref{eq13} and \eqref{coro1}, it is easy to verify that, sequences $\{p_k=e^{-k}, k=0,1, \cdots\}, \{p_k=e^{-e^{k}}, k=0,1, \cdots\}, \{p_k=e^{-e^{e^k}}, k=0,1, \cdots\}, \cdots$, satisfy Eqs. \eqref{eq13} and \eqref{coro1}. We first give a proof on why these sequences can guarantee the finite $P_{\rm avg}$.

Above sequences are all decreasing sequences with $\lim_{k \to \infty} p_k=0$. According to Corollary 1, we have to prove that Eq. \eqref{coro1} holds for these sequences. For these sequences, they obey
\begin{equation}
\lim_{k\to \infty}\frac{p_k}{\mu_k}=\lim_{k\to \infty}\frac{p_k}{1-p_k}=0. \label{thm2p1}
\end{equation}
Besides, with $k \to \infty$, $\ln(p_{k+1})$ and $\ln(p_k)$ are the same-order infinity. Thus, we obtain
\begin{equation}
\lim_{k\to \infty}\frac{\ln(p_{k+1})}{\ln(p_k)}=C, \label{thm2p2}
\end{equation}
where $C\in\mathbb{R}$ is a constant. According to Eqs. \eqref{thm2p1} and \eqref{thm2p2}, these two limits have finite values. Therefore, according to the Limit Laws \cite{rudin}, the limit of a product is equal to the product of the limits, which proves that for these sequences, 
\begin{equation}
\lim_{k \to \infty}\frac{p_k \ln(p_{k+1})}{\mu_{k}\ln(p_k)}=0.
\end{equation}

The length violation probability for the above transmission policy is given by
\begin{align}
\varepsilon(q_{\rm th})&=\sum_{k=q_{\rm th}+1}^{\infty} \pi_k \nonumber \\
&=\pi_0 \sum_{k=q_{\rm th}+1}^{\infty} \prod_{m=0}^{k-1}\bigg(\frac{p_m}{\mu_{m+1}}\bigg) \label{thm2p3}
\end{align}
Since $\mu_{m}<1$, $p_m/\mu_{m+1}<1$, and $\pi_0<1$, $\varepsilon(q_{\rm th})$ satisfies
\begin{equation}
\pi_0 \!\!\! \sum_{k=q_{\rm th}+1}^{\infty} \prod_{m=0}^{k-1}p_m < \varepsilon(q_{\rm th})< \!\!\! \sum_{k=q_{\rm th}+1}^{\infty} \prod_{m=0}^{q_{\rm th}}\bigg(\frac{p_m}{\mu_{m+1}}\bigg). \label{thm2p4}
\end{equation}

For sequences $\{p_k=e^{-k}, k=0,1, \cdots\}, \{p_k=e^{-e^{k}}, k=0,1, \cdots\}, \{p_k=e^{-e^{e^k}}, k=0,1, \cdots\}, \cdots$, $\mu_k=1-p_k\approx 1$ with $k \to \infty$. Thus, according to Eq. \eqref{thm2p4}, the decay rate of the queue length violation probability is dominated by $\sum_{k=q_{\rm th}+1}^{\infty} \prod_{m=0}^{k-1}p_m $ It indicates that, by adopting the above transmission policy, the exponent of the queue length violation probability can have an arbitrarily high decay rate with finite $P_{\rm avg}$.   
\end{IEEEproof}

In Theorem 2, we have proved that the Rayleigh fading channel belongs to Scenario 3. 
Based on the above lemmas and theorems, we will then conduct the tail distribution analysis on the more general fading distributions. In Theorem 3, we generalize the analysis to the Nakagami-m fading channel. According to \cite{gold}, the channel power gain distribution of Nakagami-m is given by
\begin{equation}
f_{|h|^2}(x)=\bigg(\frac{m}{\Omega}\bigg)^m \frac{x^{m-1}}{\Gamma(m)} e^{-\frac{mx}{\Omega}}, \quad x>0, \, m\geq 0.5,
\end{equation}
where $m$ and $\Omega$ are two parameters. Besides, $\Gamma(m)$ is the Gamma function. To simplify the derivation, we will discuss the distribution of $|h[n]|^2$ with $\Omega=1$ in Theorem 3, which is given by
\begin{equation}
g(x)=\frac{m^m}{\Gamma(m)} x^{m-1}e^{-mx} \quad x>0, \, m\geq 0.5.
\end{equation}

\textbf{Theorem 3.} For a wireless link transmission system with Nakagami-m channels,
\begin{itemize}
	\item if $m \geq 2$, it belongs to Scenario 1;
	\item if $1 \leq m <2$, it belongs to Scenario 3;
	\item if $0.5  \leq m <1$, it belongs to Scenario 2.
\end{itemize}

\begin{IEEEproof}
If $m\geq2$, we obtain
\begin{equation}
\lim_{x\to 0^+}g(x)=0,
\end{equation}
and 
\begin{align}
\lim_{x \to 0^+} g^{'}(x)=\frac{m^m}{\Gamma(m)}\bigg( \lim_{x \to 0^+}&(m-1)x^{(m-2)}e^{-mx}\nonumber \\
&-mx^{m-1}e^{-mx}\bigg)=0.
\end{align}
According to Lemma 1, we have proved that if $m\geq2$, the system can achieve zero queue length violation probability with finite $P_{\rm avg}$. Thus, with $m\geq 2$, the system belongs to Scenario 1.

For $m=1$, we have proved in Theorems 1 and 2 that the tail distribution of queue length can have arbitrary-decay-rate exponents with finite $P_{\rm avg}$. Based on this result, we then prove that for $1<m<2$, the system belongs to Scenario 3.

Given $m$, $m^m$ and $\Gamma(m)$ are finite positive constants. Intuitively, whether the system belongs to Scenario 3 should be determined by the remaining part in $g(x)$ except these constants. Thus, for the simplicity of expression in the following proof, we define $\ell(x)$ as
\begin{equation}
\ell(x)=\frac{\Gamma(m)}{m^m} g(x)=x^{m-1}e^{-mx}, \quad x>0, \, m \geq 0.5. \label{coro2p1}
\end{equation}

Here, we provide a sketch of the proof of the system belonging to Scenario 3 when $1<m<2$. First, we will prove that $\ell(x)$ with $1<m<2$ can ensure that the system belongs to Scenario 3 with the help of Theorems 1 and 2. Based on this conclusion, we will then prove $g(x)=\frac{m^m}{\Gamma(m)} \ell(x)$, $1<m<2$, can ensure that the system belongs to Scenario 3.

We see $\ell(x)$ as a function of $m$, which we denote as $w(m)$. The derivative of $w(m)$ is given by
\begin{equation}
	w^{'}(m)=e^{-mx}x^{m-1}\big(\ln x -x\big). \label{coro2p2}
\end{equation}
Since for arbitrary $x>0$, $\ln x-x<0$, we have $w'(m)<0$ with $1 \leq m <2$. Thus, $w(m)$ is a decreasing function on $[1,2)$. Then, for arbitrary $x>0$, we have $w(1)>w(m)$, $m \in (1,2)$, which is equivalent to 
\begin{equation}
	e^{-x}>x^{m-1}e^{-mx}, \quad m \in (1,2). \label{coro2p3}
\end{equation}
Let $h_{\rm th}^k(m)$ denote the channel power gain threshold of the heuristic policy under the queue length $k$ given $m$. Then, given the sequence $\{p_k,k=0,1,\cdots\}$, the relationship between $p_k$ and $h_{\rm th}^k(m)$ is given by
\begin{equation}
	p_k=\int_{0}^{h_{\rm th}^k(m)} x^{m-1}e^{-mx} dx. \label{coro2p4}
\end{equation}
By combining Eqs. \eqref{coro2p3} and \eqref{coro2p4}, we obtain that, with given $p_k$, $h_{\rm th}^k(1)<h_{\rm th}^k(m)$, $m\in(1,2)$. 

Let $C_k(m)$ denote the average power consumption under the queue length $k$ given $m\in(1,2)$. By comparing with Eq. \eqref{ck}, we find that the integrand in the definition of $C_k(m)$, i.e.,  $x^{m-2}e^{-mx}$, is smaller than $e^{-x}/x$ according to Eq. \eqref{coro2p3}. Besides, the integral interval in the definition of $C_k(m)$ is $(h_{\rm th}^k(m), \infty)$. The length of $(h_{\rm th}^k(m), \infty)$ is smaller than $(h_{\rm th}^k(1), \infty)$ since $h_{\rm th}^k(1)<h_{\rm th}^k(m)$, $m\in(1,2)$. Therefore, given the sequence $\{p_k, k=0,1,\cdots\}$, with smaller integrand and integral interval, 
\begin{equation}
	C_k(m)<C_k(1), \quad m\in(1,2). \label{coro2p5}
\end{equation}

Note that, the steady state probability of queue length under the heuristic policy is determined by the $\{p_k, k=0,1,\cdots\}$ sequence only. Thus, given the same $\{p_k, k=0,1,\cdots\}$ sequence, the steady state probability of queue length with $m=1$ is equal to that with $m\in (1,2)$. Moreover, for each queue length $k$, Eq. \eqref{coro2p5} holds. Thus, we have proved that, with the same sequence $\{p_k, k=0,1,\cdots\}$, $P_{\rm avg}$ of $m=1$ is larger than $P_{\rm avg}$ of $m\in (1,2)$, which indicates $\ell(x)$ with $1<m<2$ can ensure that the system belongs to Scenario 3.

Then, we present the proof that $g(x)=\frac{m^m}{\Gamma(m)} \ell(x)$, $1<m<2$, can ensure that the system belongs to Scenario 3.

Let $h^{k,1}_{\rm th}(m)$ denote the channel power gain threshold of the heuristic policy under the queue length $k$ given $m$ for $\ell(x)$, while $h^{k,2}_{\rm th}(m)$ denote that for $g(x)$. The relationships between $p_k$ and $h^{k,1}_{\rm th}(m)$, $h^{k,2}_{\rm th}(m)$ are given by
\begin{equation}
	p_k=\int_{0}^{h_{\rm th}^{k,1}(m)} x^{m-1}e^{-mx} dx,
\end{equation}
\begin{equation}
	p_k=\int_{0}^{h_{\rm th}^{k,2}(m)} \frac{m^m}{\Gamma(m)}x^{m-1}e^{-mx} dx. 
\end{equation}
Since $m^m/\Gamma(m)>1$ when $m\in(1,2)$, we obtain $h^{k,1}_{\rm th}(m)>h^{k,2}_{\rm th}(m)$. Let $C_k^{(1)}(m)$ and $C_k^{(2)}(m)$ denote the average power consumption under the queue length $k$ given $m$ of $\ell(x)$ and $g(x)$, respectively. $C_k^{(1)}(m)$ and $C_k^{(2)}(m)$ are given by
\begin{equation}
	C_k^{(1)}(m)=a_k\int_{h^{k,1}_{\rm th}(m)}^{\infty} x^{m-2}e^{-mx}dx, \label{coro2p6}
\end{equation}
\begin{equation}
	C_k^{(2)}(m)=\frac{a_k m^m}{\Gamma(m)}\int_{h^{k,2}_{\rm th}(m)}^{\infty} x^{m-2}e^{-mx}dx, \label{coro2p7}
\end{equation}
where $a_k$ is given by
\begin{equation}
	a_k=
	\begin{cases}
		e^2-1, \quad &k\geq 1, \\
		e-1, \quad &k=0.
	\end{cases}
\end{equation}
Based on Eq. \eqref{coro2p6} and $h^{k,1}_{\rm th}(m)>h^{k,2}_{\rm th}(m)$, we decompose Eq. \eqref{coro2p7} as 
\begin{align}
	C_k^{(2)}(m)=&\frac{a_k m^m}{\Gamma(m)} \int_{h^{k,2}_{\rm th}(m)}^{h^{k,1}_{\rm th}(m)} x^{m-2}e^{-mx}dx \nonumber \\
	&+ \frac{a_k m^m}{\Gamma(m)}\int_{h^{k,1}_{\rm th}(m)}^{\infty} x^{m-2}e^{-mx}dx, \label{coro2p8}
\end{align}
where the second term of the right side of Eq. \eqref{coro2p8} is equal to $\frac{m^m}{\Gamma(m)}C_k^{(1)}(m)$. Let $P_{\rm avg}^{(1)}$ and $P_{\rm avg}^{(2)}$ denote the average power consumption of the heuristic policy with $\ell(x)$ and $g(x)$, respectively. Then, based on Eq. \eqref{coro2p8}, $P_{\rm avg}^{(2)}$ is given by
\begin{equation}
	P_{\rm avg}^{(2)}=\sum_{k=0}^{\infty}\bigg(\pi_k \frac{a_k m^m}{\Gamma(m)} \int_{h^{k,2}_{\rm th}(m)}^{h^{k,1}_{\rm th}(m)} x^{m-2}e^{-mx}dx\bigg) +\frac{m^m}{\Gamma(m)}P_{\rm avg}^{(1)}. \label{coro2p9}
\end{equation}

As we proved before, $P_{\rm avg}^{(1)}$ is finite with $\{p_k, k=0,1,\cdots\}$ satisfying Eq. \eqref{eq13}. Thus, $\frac{m^m}{\Gamma(m)}P_{\rm avg}^{(1)}$ is finite with $m \in (1,2)$ with  the same $\{p_k, k=0,1,\cdots\}$ sequence. 

For the first term on the right side of Eq. \eqref{coro2p8}, it satisfies
\begin{align}
	\frac{a_km^m}{\Gamma(m)} \int_{h^{k,2}_{\rm th}(m)}^{h^{k,1}_{\rm th}(m)} x^{m-2}e^{-mx}dx&<\frac{a_km^m}{\Gamma(m)} \int_{h^{k,2}_{\rm th}(m)}^{h^{k,1}_{\rm th}(m)} x^{m-2}dx \nonumber \\
	&<\frac{a_km^m}{\Gamma(m)} \int_{0}^{h^{k,1}_{\rm th}(m)} x^{m-2}dx \nonumber \\
	&=\frac{a_km^m}{\Gamma(m)} \underbrace{\frac{\Big(h^{k,1}_{\rm th}(m)\Big)^{m-1}}{m-1}}_{r_k(m)}.
\end{align}

For $m\in(1,2)$, given a decreasing sequence $\{p_k, k=0,1,\cdots\}$ which satisfies $\lim_{k \to \infty}p_k=0 $, we obtain that $\lim_{k \to \infty} r_k(m)=0$, $m \in (1,2)$. Thus, given sequence $\{p_k, k=0,1,\cdots\}$ satisfying Eq. \eqref{coro1}, the first term on the right side of Eq. \eqref{coro2p8} is finite. Now we have proved that given sequence $\{p_k, k=0,1,\cdots\}$ satisfying Eq. \eqref{eq13}, $P_{\rm avg}^{(2)}$ is finite, which indicates that $g(x)$ with $1<m<2$ can ensure that the system belongs to Scenario 3.

For $0.5\leq m<1$, we find that 
\begin{equation}
	\lim_{\epsilon \to 0^+}g(x)= \infty.
\end{equation}
Besides, similar to the proof for $m \in (1,2)$, given the sequence $\{p_k,k=0,1,\cdots\}$, $P_{\rm avg}$ of $m=1$ is smaller than $P_{\rm avg}$ of $m \in [0.5,1)$. Therefore, we can not ensure whether the system belongs to Scenario 1 or Scenario 3 when $m\in[0.5,1)$. However, at least we can guarantee that the system belongs to Scenario 2 by adopting the fixed power policy. 
\end{IEEEproof}

\section{Simulation Result}
In this section, we conduct simulations to verify the proposed conclusion in Section III.
\begin{figure}[t]
	\centerline{\includegraphics[width=8.5cm]{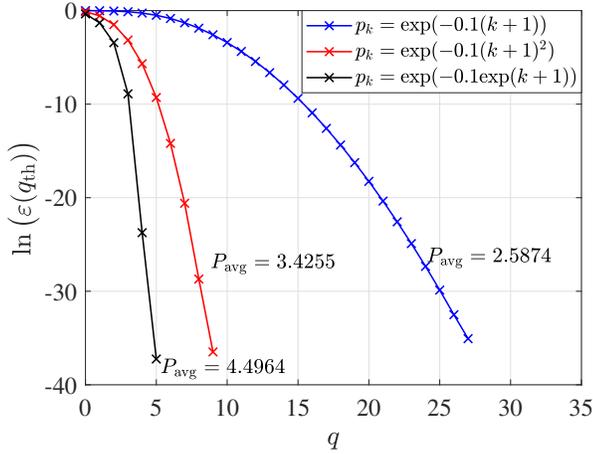}}
	\caption{Exponent of the queue length tail distribution with different $p_k$ sequence in the Rayleigh fading channel.}
	\label{fig1}
\end{figure}
\begin{figure}[t]
	\centerline{\includegraphics[width=8.5cm]{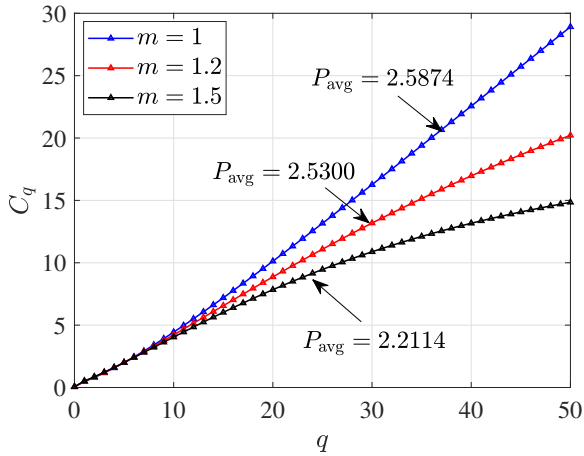}}
	\caption{$C_q$ versus $q$ with the $p_k=\exp(-0.1(k+1))$ in Nakagami-m channels.}
	\label{fig2}
\end{figure}

In Fig. \ref{fig1}, we choose three $\{p_k, k=0,1,\cdots\}$ sequences of the transmission policy mentioned in Section III in the Rayleigh fading channel. In Section II we assume the system is equipped with an  infinite-length buffer, while in the simulation we set the buffer size equal to 50 to approximate the result. The other parameters are set as $A=1$, $\mathbb{E}\{|h[n]|^2\}=1$, and $\sigma^2=1$. As shown in Fig. \ref{fig1}, the average power consumption of the transmission policy with these three $p_k$ sequences are 2.5874, 3.4255, and 4.4964, respectively. Besides, the queue length tail distribution with $p_k=\exp(-0.1\exp(k+1))$ has the maximum-decreasing-rate exponent. These results validate Theorems 1 and 2.  Fig. \ref{fig1} also indicates that, in the Rayleigh fading channel, the queue length tail distribution can have an exponent which decreases faster than the linear rate proposed in LDT with the finite average power consumption.

In Fig. \ref{fig2}, we compare the $C_q$, i.e., the average power consumption with queue length $q$, in different Nakagami-m channels by adopting the transmission policy mentioned in Section III with $1 \leq m <2$. Since the $p_k$ sequence is same, the steady state probabilities for these three systems with $m=1$, $m=1.2$, and $m=1.5$ are identical. As shown in Fig. \ref{fig2}, with a higher $m$, the $C_q$ is less, which results in the less $P_{\rm avg}$. This tradeoff meets our expectations since higher $m$ means better performance on avoiding severe fading.

\section{Conclusion}
In this paper, we analyzed the decay rate of the queue length tail distribution with the cross-layer design. Specifically, we divided the communication systems into three scenarios according to the decay rate of the queue length tail distribution with the finite average power consumption. We presented sufficient conditions for the systems belonging to Scenario 1 and Scenario 2, respectively. Then, we conceived a transmission policy to analyze the communication system with Rayleigh fading channels. By presenting the sufficient condition that the arbitrary-decay-rate tail distribution with finite average power consumption exists, we proved that the system with Rayleigh fading channels belongs to Scenario 3. Finally, the analysis for Rayleigh fading channels was generalized to Nakagami-m fading channels. The analysis in this paper provides the achievable decay rate of the queue length tail distribution, which can be used to instruct the estimate and analysis for the optimal cross-layer control policy. Future promising research directions include generalizing the analysis for more complex channel models and finding necessary and sufficient conditions for the systems belonging to three scenarios, respectively.

\bibliographystyle{IEEEtran}
\begin{spacing}{1.03}
	\bibliography{mybib}

\begin{thebibliography}{10}
\providecommand{\url}[1]{#1}
\csname url@samestyle\endcsname
\providecommand{\newblock}{\relax}
\providecommand{\bibinfo}[2]{#2}
\providecommand{\BIBentrySTDinterwordspacing}{\spaceskip=0pt\relax}
\providecommand{\BIBentryALTinterwordstretchfactor}{4}
\providecommand{\BIBentryALTinterwordspacing}{\spaceskip=\fontdimen2\font plus
\BIBentryALTinterwordstretchfactor\fontdimen3\font minus
  \fontdimen4\font\relax}
\providecommand{\BIBforeignlanguage}[2]{{%
\expandafter\ifx\csname l@#1\endcsname\relax
\typeout{** WARNING: IEEEtran.bst: No hyphenation pattern has been}%
\typeout{** loaded for the language `#1'. Using the pattern for}%
\typeout{** the default language instead.}%
\else
\language=\csname l@#1\endcsname
\fi
#2}}
\providecommand{\BIBdecl}{\relax}
\BIBdecl

\bibitem{6G}
K.~B. {Letaief}, W.~{Chen}, Y.~{Shi}, J.~{Zhang}, and Y.~A. {Zhang}, ``The
  roadmap to {6G: AI} empowered wireless networks,'' \emph{IEEE Commun. Mag.},
  vol.~57, no.~8, pp. 84--90, Aug. 2019.

\bibitem{tsn}
A.~Nasrallah, A.~S. Thyagaturu, Z.~Alharbi, C.~Wang, X.~Shao, M.~Reisslein, and
  H.~ElBakoury, ``Ultra-low latency ({ULL}) networks: The {IEEE TSN and IETF
  DetNet} standards and related {5G ULL} research,'' \emph{IEEE Commun. Surveys
  Tuts.}, vol.~21, no.~1, pp. 88--145, Sept. 2019.

\bibitem{chen}
X.~Chen, W.~Chen, J.~Lee, and N.~B. Shroff, ``Delay-optimal buffer-aware
  scheduling with adaptive transmission,'' \emph{IEEE Trans. Commun.}, vol.~65,
  no.~7, pp. 2917--2930, Apr. 2017.

\bibitem{wang}
M.~Wang, J.~Liu, W.~Chen, and A.~Ephremides, ``Joint queue-aware and
  channel-aware delay optimal scheduling of arbitrarily bursty traffic over
  multi-state time-varying channels,'' \emph{IEEE Trans. Commun.}, vol.~67,
  no.~1, pp. 503--517, Oct. 2019.

\bibitem{zhao}
X.~Zhao, W.~Chen, J.~Lee, and N.~B. Shroff, ``Delay-optimal and
  energy-efficient communications with markovian arrivals,'' \emph{IEEE Trans.
  Commun.}, vol.~68, no.~3, pp. 1508--1523, Dec. 2020.

\bibitem{yang}
J.~Yang and S.~Ulukus, ``Delay-minimal transmission for average power
  constrained multi-access communications,'' \emph{IEEE Trans. Wireless
  Commun.}, vol.~9, no.~9, pp. 2754--2767, Jul. 2010.

\bibitem{ec}
{D. Wu} and R.~{Negi}, ``Effective capacity: a wireless link model for support
  of quality of service,'' \emph{IEEE Trans. Wireless Commun.}, vol.~2, no.~4,
  pp. 630--643, Jul. 2003.

\bibitem{ecqos}
S.~Shakkottai, ``Effective capacity and {QoS} for wireless scheduling,''
  \emph{IEEE Trans. Autom. Control}, vol.~53, no.~3, pp. 749--761, Apr. 2008.

\bibitem{spldt}
L.~Ying, R.~Srikant, A.~Eryilmaz, and G.~Dullerud, ``A large deviations
  analysis of scheduling in wireless networks,'' \emph{IEEE Trans. Inf.
  Theory}, vol.~52, no.~11, pp. 5088--5098, Oct. 2006.

\bibitem{lya}
V.~J. Venkataramanan and X.~Lin, ``On the queue-overflow probability of
  wireless systems: A new approach combining large deviations with lyapunov
  functions,'' \emph{IEEE Trans. Inf. Theory}, vol.~59, no.~10, pp. 6367--6392,
  Jun. 2013.

\bibitem{inversion}
A.~J. Goldsmith and P.~P. Varaiya, ``Capacity of fading channels with channel
  side information,'' \emph{IEEE Trans. Inf. Theory}, vol.~43, no.~6, pp.
  1986--1992, Nov. 1997.

\bibitem{e1}
M.~Abramowitz and I.~A. Stegun, \emph{Handbook of mathematical functions with
  formulas, graphs, and mathematical tables}.\hskip 1em plus 0.5em minus
  0.4em\relax US Government printing office, 1964, vol.~55.

\bibitem{rudin}
W.~Rudin \emph{et~al.}, \emph{Principles of mathematical analysis}.\hskip 1em
  plus 0.5em minus 0.4em\relax McGraw-hill New York, 1964, vol.~3.

\bibitem{gold}
A.~Goldsmith, \emph{Wireless communications}.\hskip 1em plus 0.5em minus
  0.4em\relax Cambridge university press, 2005.

\bibitem{online}
{L. Li, W. Chen, and K. B. {Letaief}, "Bounding Queue Length Violation
  Probability of Joint Channel and Buffer Aware Transmission," 2021,
  \emph{arXiv:2111.06569}. [Online]. Available:
  \url{https://arxiv.org/abs/2111.06569}}.

\end{thebibliography}
\end{spacing}

\end{document}